\documentclass[twocolumn,showpacs,superscriptaddress]{revtex4-1} 

\usepackage{amssymb}
\usepackage{subfigure}
\usepackage{amsmath}

\usepackage{graphicx}
\usepackage{epstopdf}
\usepackage{color}
\usepackage{ulem}

\newcommand{\scl}{0.2} 

\begin{document}

\title{Antiresonances and Ultrafast Resonances in a Twin Photonic Oscillator }

\date{\today}

\author{Yannis Kominis}
\affiliation{School of Applied Mathematical and Physical Science, National Technical University of Athens, Athens, Greece	}

\author{Kent D. Choquette}
\affiliation{Department of Electrical and Computer Engineering, University of Illinois, Urbana, Illinois, USA}

\author{Anastasios Bountis}
\affiliation{Department of Mathematics, School of Science and Technology, Nazarbayev University, Astana, Republic of Kazakhstan}

\author{Vassilios Kovanis}
\affiliation{Department of Physics, School of Science and Technology, Nazarbayev University, Astana, Republic of Kazakhstan}

\begin{abstract}
We consider the properties of the small-signal modulation response of symmetry-breaking phase-locked states of twin coupled semiconductor lasers. The extended stability and the varying asymmetry of these modes allows for the introduction of a rich set of interesting modulation response features, such as sharp resonances and anti-resonance as well as efficient modulation at very high frequencies exceeding the free running relaxation frequencies by orders of magnitude. 
\end{abstract}

\maketitle

\section{Introduction}
The optical coupling of two or more semiconductor lasers and its effects on the output of such systems have been intensively studied for more than four decades \cite{Spencer_72} both theoretically and experimentally. The coupling introduces a rich set of complex dynamical features in the system such as phase locking, instabilities, bifurcations and limit cycles  \cite{Wang_88, Winful&Wang_88, Winful_90, Kovanis_97, Winful_92}. From the technological point of view, systems of coupled semiconductor lasers offer capabilities for numerous integrated photonics applications \cite{Coldren_book} as transmitters in high-speed optical communications and optical interconnects, high-power laser sources, tunable photonic oscillators, controllable optical beam shaping and steering elements and ultrasensitive sensors \cite{Tatum_15, Choquette_13, Choquette_15,  Choquette_17, Choquette_18}.  The underlying model of such structures is a system of coupled rate equations governing the time evolution of the electric fields amplitudes and phases as well as the carrier dynamics, with the carrier-induced nonlinearity playing a crucial role due to a non-zero linewidth enhancement factor describing an amplitude-phase coupling mechanism \cite{Erneux_book}.\
For most of the important technological applications, the existence of stable phase-locked modes is crucial for the coherent emission of optical power. In addition, the functionality of the system depends crudially on the capability of modulating its output by varying injection currents \cite{Acebo_02, Grasso_06, Fryslie_15, Xiao_17, Zhu_18}. For single semiconductor lasers the upper limit for efficient modulation is restricted by its relaxation oscillation frequency that typically cannot exceed a few GHz \cite{Yariv_85}. An increased electrical pumping rate increases the relaxation oscillation frequency where the resonance takes place but cannot control the profile of the modulation response. More complex configurations based on optical injection schemes have also been considered for the control of the modulation response of the system \cite{Simpson_96, Simpson_97, Varangis_97, Wu_07, Wu_08a, Wu_08b, Lester_14}. However, these configurations neccesitate the utilization of an optical isolator which is quite challenging for photonic integration \cite{Christodoulides_17}. The simplest system of coupled semiconductor lasers, that can be used in integrated photonic circuits, consists of a pair of twin lasers with equal pumping. It is known that out-of-phase modulation of the two lasers can result in effective modulation beyond the relaxation oscillation frequency \cite{Agrawal_85, WilsonDeFreez_91a, WilsonDeFreez_91b}. However, the fact that the stability of the modulated phase locked state is restricted in the parameter space of the system may render such results impractical for realistic applications \cite{Golden_08}. In a recent work, we have uncovered the existence of a new symmetry-breaking phase-locked state that is stable for a much larger region of the parameter space for the whole range of values of the optical coupling coefficient \cite{Kominis_17a, Kominis_17b}, in comparison to the well-known symmetric states \cite{Winful&Wang_88}. \

In this paper, we explore the small signal modulation response of the system, when these stable phase-locked states of varying asymmetry are modulated. We show that the extended stability along with the asymmetric character of these states, results in a frequency response with a rich set of qualitatively different features, including the asymmetry of the amplitude response of the two lasers, the existence of sharp resonances and anti-resonances as well as the capability of efficient stable modulation of the system at extremely high frequencies that are orders of magnitudes larger than its free running relaxation frequency.   

\section{Rate equation model and symmetry-breaking phase locking}
The dynamics of an array of two indentical evanescently coupled semiconductor lasers is governed by the following coupled single-mode rate equations for the amplitude of the normalized electric fields $X_i$, the phase difference $\theta$ between the fields and the normalized excess carrier density $Z_i$ of each semiconductor laser:
\begin{eqnarray}
 \frac{dX_1}{dt}&=&X_1Z_1-\Lambda X_2\sin\theta \nonumber \\
 \frac{dX_2}{dt}&=&X_2Z_2+\Lambda X_1\sin\theta \nonumber \\
 \frac{d\theta}{dt}&=& \alpha(Z_1-Z_2)+\Lambda\left(\frac{X_1}{X_2}-\frac{X_2}{X_1}\right)\cos\theta   \label{pair} \\
T\frac{dZ_1}{dt}&=&P_1- Z_1-(1+2 Z_1)X_1^2 \nonumber \\
T\frac{dZ_2}{dt}&=&P_2- Z_2-(1+2 Z_2)X_2^2 \nonumber
\end{eqnarray}
Here $\alpha$ is the linewidth enhancement factor, $\Lambda$ is the normalized coupling constant, $P_{1,2}$ are the normalized excess pumping rates, $T$ is the ratio of carrier to photon lifetimes, and $t$ is the time normalized to the photon lifetime $\tau_p$ \cite{Winful&Wang_88, Choquette_13}. In the following, we consider parameter values corresponding to recent experiments on coherently coupled phased photonic crystal vertical cavity lasers \cite{Choquette_17mod} as shown in Table I.

\begin{table}
\centering
 \begin{tabular}{|c c c |} 
 \hline
Symbol & Parameter & Value  \\ 
 \hline \hline
 $\alpha$ & Linewidth enhancement factor & $4$  \\ 
 \hline
 $\tau_c$ & Carrier lifetime & $2$ns \\
\hline
 $\tau_p$ & Photon lifetime & $2 \times 10^{-3}$ns \\
 \hline
 $T$ & Ratio of carrier to photon lifetime & $1000$  \\ 
 \hline
 $\Lambda$ & Normalized coupling coefficient & $10^{-4} - 10^0$ \\ 
 \hline
\end{tabular}
\caption{Realistic parameter values for coherently coupled phased photonic crystal vertical cavity lasers \cite{Choquette_17mod}.}
\end{table}

Under equal pumping $(P_1=P_2=P_0)$, the phase-locked states of the system (\ref{pair}), are given by setting the time derivatives of the system equal to zero and their stability is determined by the eigenvalues of the Jacobian of the linearized system. The in-phase and out-of-phase phase-locked modes corresponding to $X_1=X_2=\sqrt{P_0}$, $Z_1=Z_2=0$ and $\theta=0,\pi$ are known to be stable for the strong ($\Lambda>\alpha P_0 /(1+2P_0)$) and the weak ($\Lambda<(1+2P_0)/2\alpha T$) coupling regime, respectively \cite{Winful&Wang_88}. A symmetry-breaking phase-locked state has been shown \cite{Kominis_17a, Kominis_17b} to exist and to be stable for a large extent of coupling coefficient values, as illustrated in Fig. 1(a), with nonunitary amplitude ratio $\rho \equiv X_2/X_1$ and
\begin{equation}
\theta= \tan^{-1}\left[\frac{1}{\alpha} \frac{\rho^2-1}{\rho^2+1}\right] \label{theta_rho} \\ 
\end{equation}
\begin{equation}
X_1^2 = \frac{ \Lambda \sin\theta (\rho^2+1)}{\rho\left[(\rho^2-1)-4 \Lambda \rho \sin\theta\right]} \label{X1_rho}
\end{equation}
\begin{eqnarray}
Z_1&=&\Lambda \rho \sin\theta \nonumber \\
Z_2&=&-\frac{\Lambda}{\rho}\sin\theta \label{Z_eq}
\end{eqnarray} 
The pumping rate for this asymmetric phase-locked state is fixed at the value 
\begin{equation}
 P_0=X_1^2+(1+2X_1^2) \Lambda \rho \sin\theta \label{P0}
\end{equation}
A characteristic reference frequency of the system of the coupled twin lasers is the free running  relaxation frequency of a single laser which is pumped with a pumping value $P_0$ corresponding to the asymmetric phase-locked state (\ref{P0}) that is given by
\begin{equation}
 \Omega_{rel}=\sqrt{\frac{2P_0}{T}}
\end{equation}
Since $P_0$ depends on the coupling coefficient ($\Lambda$) and the asymmetry ($\rho$) of the repective phase-locked state, the corresponding free-running relaxation frequency varies as shown in Fig. 1(b) for $\Lambda=10^{-3}, 10^{-2}, 10^{-1}, 10^{-0.2}$, with the variation with respect to $\rho$ being more pronounced for the highly asymmetric states in the weak coupling regime.\

\begin{figure}[pt]
  \begin{center}
  \subfigure[]{\scalebox{\scl}{\includegraphics{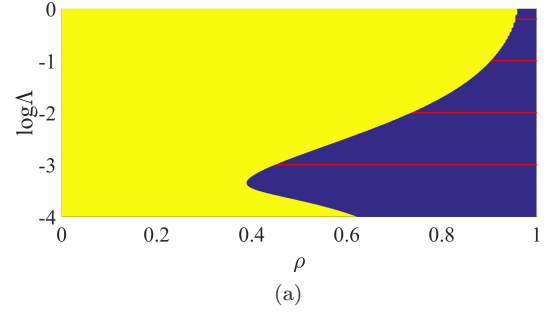}}}\\
  \subfigure[]{\scalebox{\scl}{\includegraphics{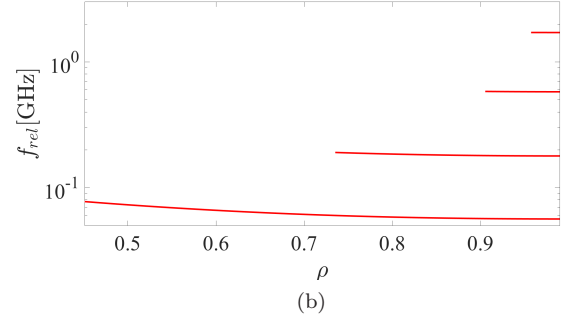}}}
  \caption{(a) Stability region (blue shaded area) of symmetry-breaking phase-locked states with electric field amplitude ratio $\rho$ for varying normalized coupling coefficient $\Lambda$. The horizontal (red) lines correspond to $\Lambda=10^{-3}, 10^{-2}, 10^{-1}, 10^{-0.2}$. (b) Free running relaxation frequency ($f_{rel}$) as a function of $\rho$ for the aforementioned values of $\Lambda$.}
  \end{center}
\end{figure}

It is worth emphasizing that the symmetry-breaking phase-locked states are stable for a much larger range of values of the coupling coefficients in comparison to the in-phase and the out-of-phase phase-locked states. The instability of the latter, has been a long standing issue preventing the observation of the direct current modulation response in such systems \cite{WilsonDeFreez_91a, WilsonDeFreez_91b, Golden_08, Choquette_17mod}.

\section{Linear modulation response}
In the following we take advantage of the explicit knowledge of the stability domain in the parameter space as well as the additional freedom to modulate asymmetric phase-locked state in order to investigate interesting features of the modulation response of this system. The small signal (linear) modulation response to a time-varying current 
\begin{eqnarray}
 P_1&=&P_0 +  \Re  \{ \delta P e^{i\omega t}\} \nonumber \\
 P_2&=&P_0 + s \Re \{ \delta P e^{i\omega t}\}
\end{eqnarray}
is calculated by linearizing the system around a stable phase-locked state, with $s=\pm 1$ corresponding to cases of in-phase and out-of-phase modulation. 
By defining $\vec{X} \equiv (X_1,X_2,\theta,Z_1, Z_2)$ and writting $\vec{X}=\vec{X}_0+\vec{\delta X}$ where $\vec{X_0}$ denotes the phase-locked state and $\vec{\delta X}$ being the small signal response of the system, the frequency response is obtained as 
\begin{equation}
\vec{\delta X}=\mathbf{H}(\omega)\vec{A} \delta P
\end{equation}
with $\vec{A}=(0, 0, 0, 1/T, s/T)$ and
\begin{equation}
\mathbf{H}(\omega)=\left(i\omega\mathbf{I}-\mathbf{J} \right)^{-1}
\end{equation}
being the transfer matrix of the linear system and $\mathbf{J}$ its Jacobian. Therefore, the electric field amplitude response is given by
\begin{equation}
 \left| \frac{\delta X_i}{\delta P}  \right| =\frac{1}{T}\left| H_{i,4}(\omega) + s H_{i,5}(\omega) \right|, \hspace{2em} i=1,2 
\end{equation}
Depending on the asymmetry of the modulated phase-locked state and the value of the coupling coefficient, the linear modulation response of the system has a rich set of qualitative different features including asymmetric frequency response of the two lasers, resonances and anti-resonances between the two lasers as well as  singificant modulation response at frequencies far beyond the free running relaxation frequency.  

\begin{figure}[pt]
  \begin{center}
  {\scalebox{\scl}{\includegraphics{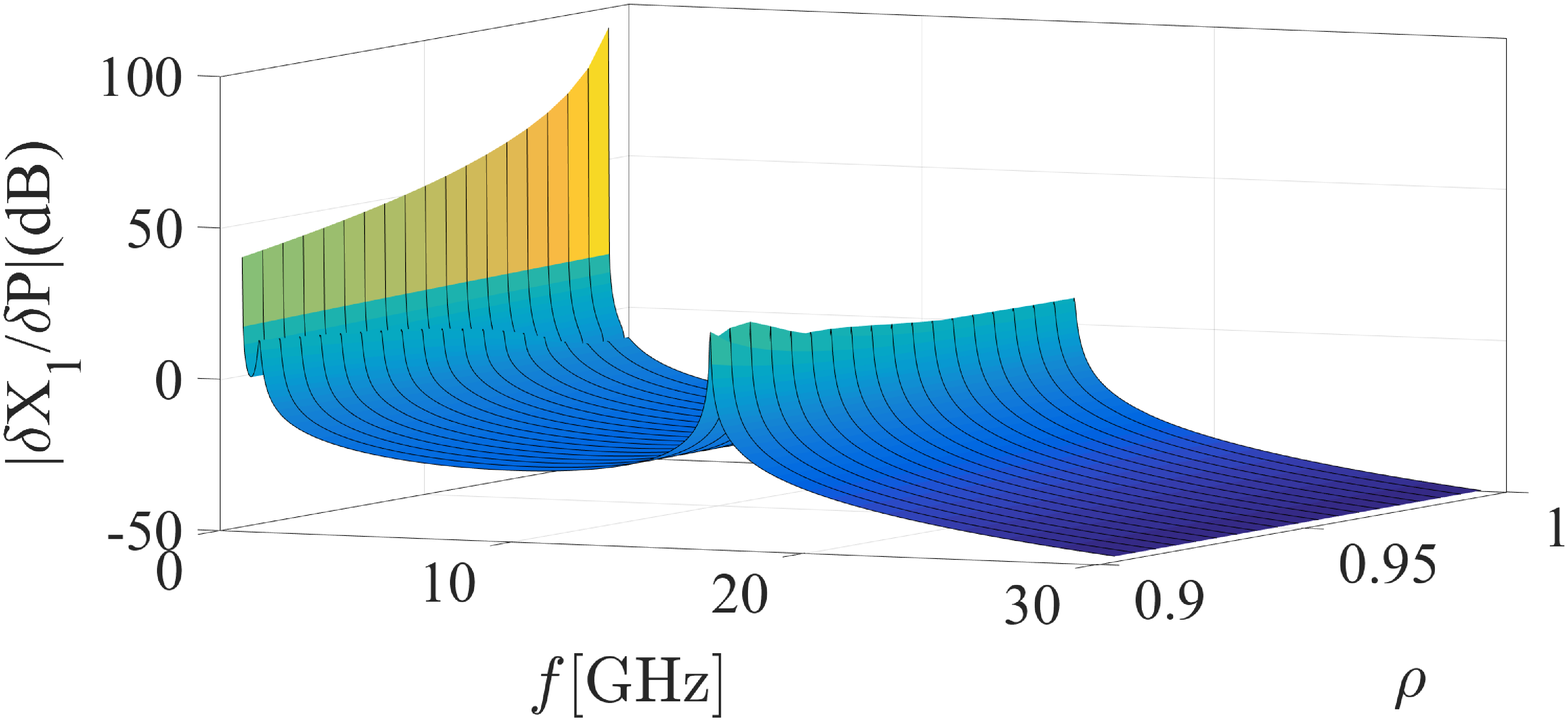}}}\\
  {\scalebox{\scl}{\includegraphics{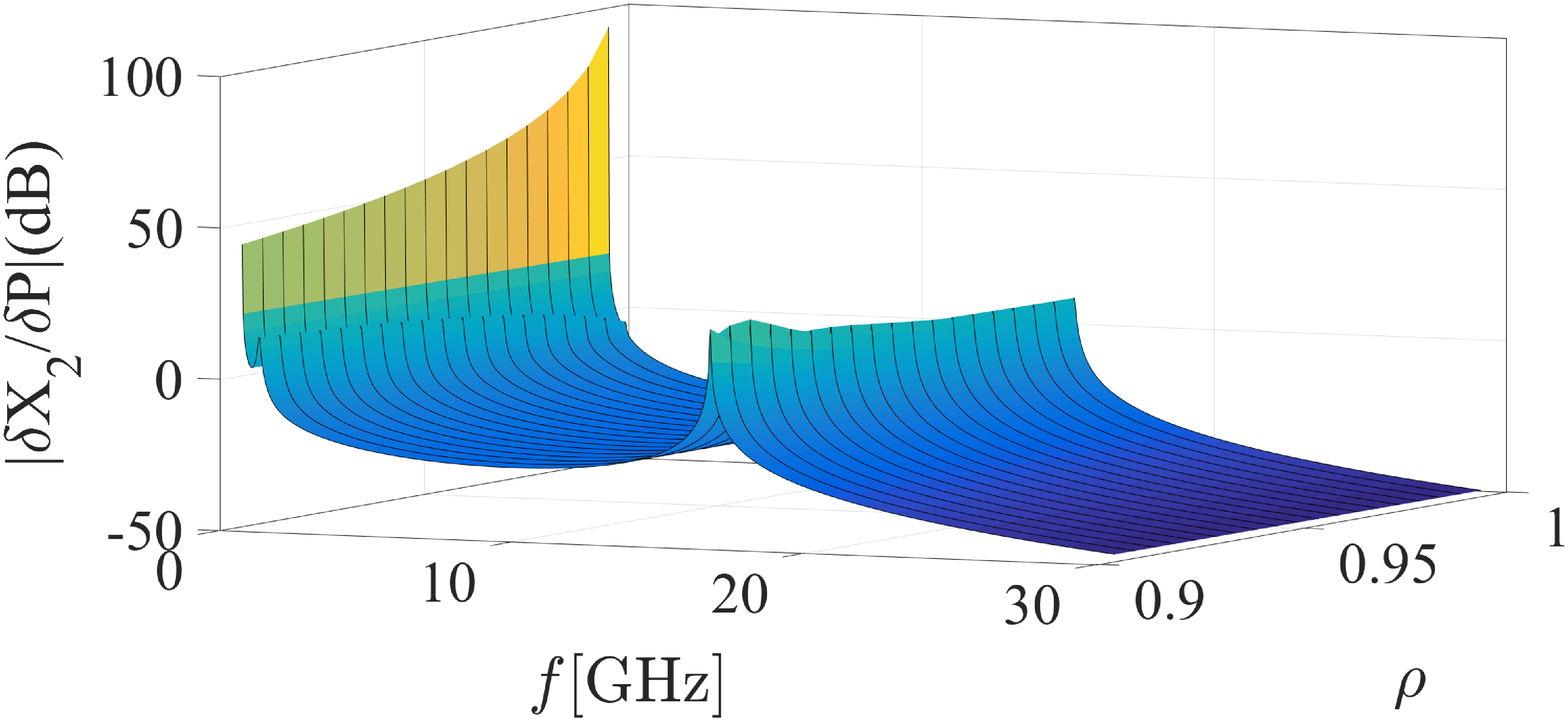}}}
  \caption{Modulation response as a function of the electric field amplitude ratio $\rho$ for $\Lambda=10^{-1}$ under out-of-phase modulation. The amplitude response is peaked at much higher frequency than the free running relaxation frequency and weakly depends on $\rho$. }
  \end{center}
\end{figure}

\begin{figure}[h!]
  \begin{center}
  {\scalebox{\scl}{\includegraphics{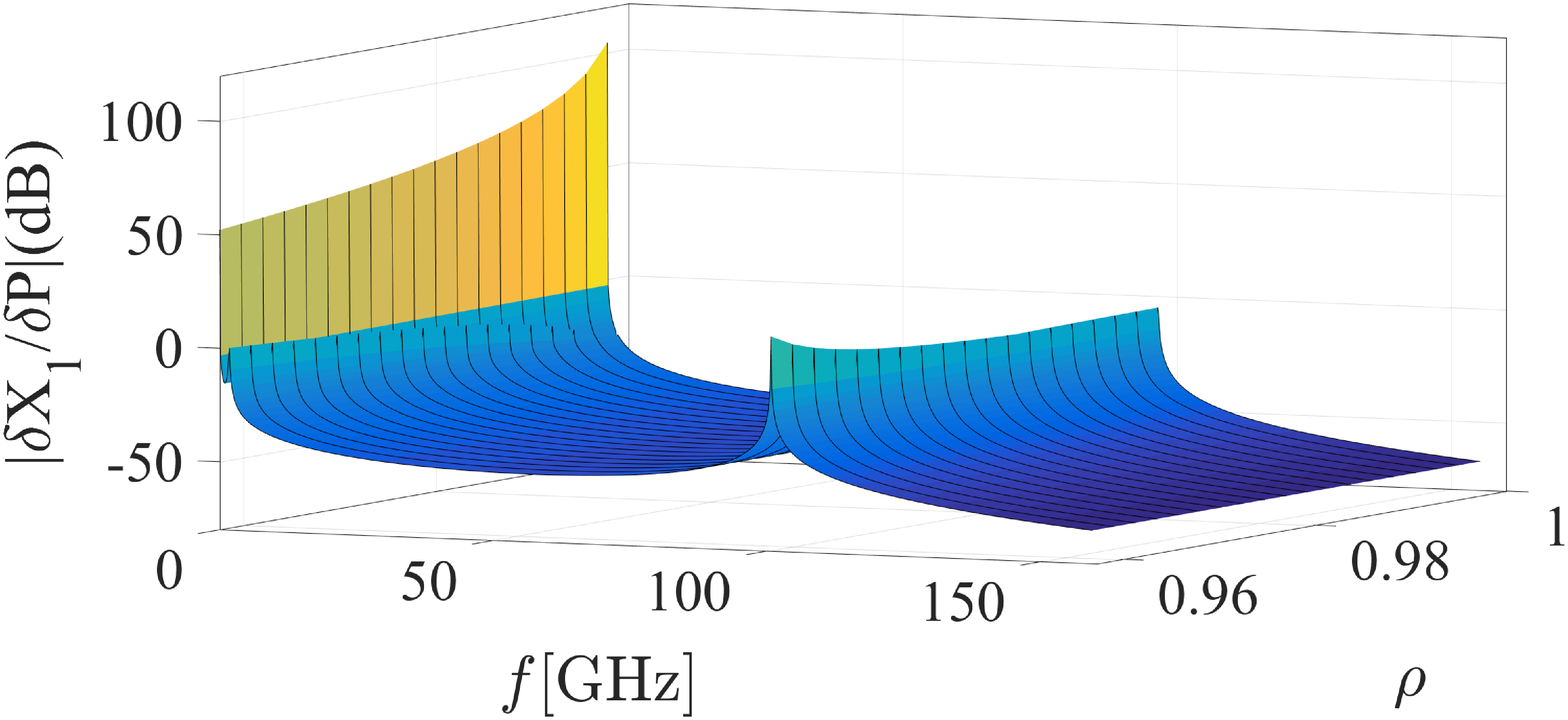}}}\\
  {\scalebox{\scl}{\includegraphics{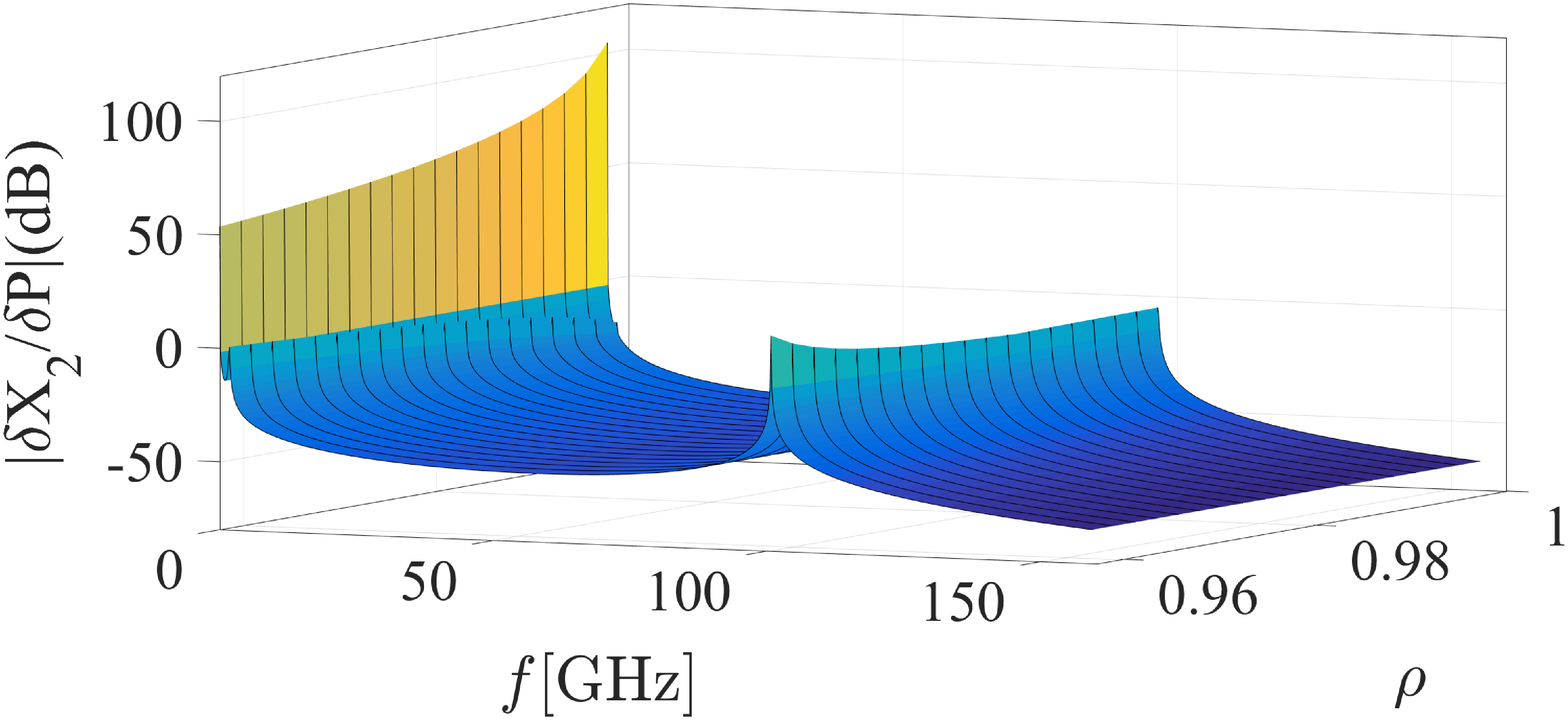}}}
  \caption{Modulation response as a function of the electric field amplitude ratio $\rho$ for $\Lambda=10^{-0.2}$ under out-of-phase modulation. The amplitude response is peaked at much higher frequency (beyond 100\,GHz) than the free running relaxation frequency and weakly depends on $\rho$. }
  \end{center}
\end{figure}

\subsection{Strong coupling}
In the strong coupling regime, the stable phase-locked states are slightly asymmetric ($\rho$ close to unity) in comparison to the weak coupling regime and the respective free running relaxation frequencies are at the order of GHz as shown in Fig. 1. In such cases, the in-phase modulation response has a peak close to the free running relaxation frequency which is smaller than 1\,GHz, for a coupling coefficient $\Lambda=10^{-1}$ (as shown in Fig. 1(b)). For the out-of-phase modulation, a sharp peak appears at a frequency that is several times larger than the free running relaxation frequency and its maximum depends weakly on the asymmetry ($\rho$) of the modulated phase-locked state; moreover the modulation response is flattened between the two peaks, as shown in Fig. 2. For even larger values of the coupling coefficient, such as $\Lambda=10^{-0.2}$ and out-of-phase modulation we have the same qualitative features, but with a resonant peak appearing at a frequency beyond 100\,GHz, which is more than 50 times larger than the free running relaxation frequency \cite{CLEO_18}, as shown in Fig. 3. As the stable phase-locked states for strong coupling are weakly asymmetric, with $\rho$ close to unity, the high-frequency peaks of the out-of-phase modulation response appear at frequencies that are close to the analytically known frequencies \cite{WilsonDeFreez_91a, WilsonDeFreez_91b}.
\begin{equation}
 \Omega=2\Lambda+\Omega_{rel}^2/\Lambda
\end{equation}
Apart from the aforementioned high-frequency peaks and the low-frequency peaks appearing at the free-running relaxation frequencies, the modulation response is peaked at the zero frequency. As a result, the modulation response drops sharply and rolls-off close to DC before being enhanced by the high-frequnecy resonance, in contrast to single laser modulation response that is typically flat or rising from the DC to the resonance \cite{Erneux_book, Coldren_book}. This effect, that is also typical for strong optical injection configurations \cite{Wu_07,Wu_08a}, can be restrictive for some high-bandwidth requiring applications and can be mitigated by the utilization of appropriate design parameter values \cite{Wu_08b} or frequency modulation \cite{Coldren_book}.    
 
\begin{figure}[pt]
  \begin{center}
  {\scalebox{\scl}{\includegraphics{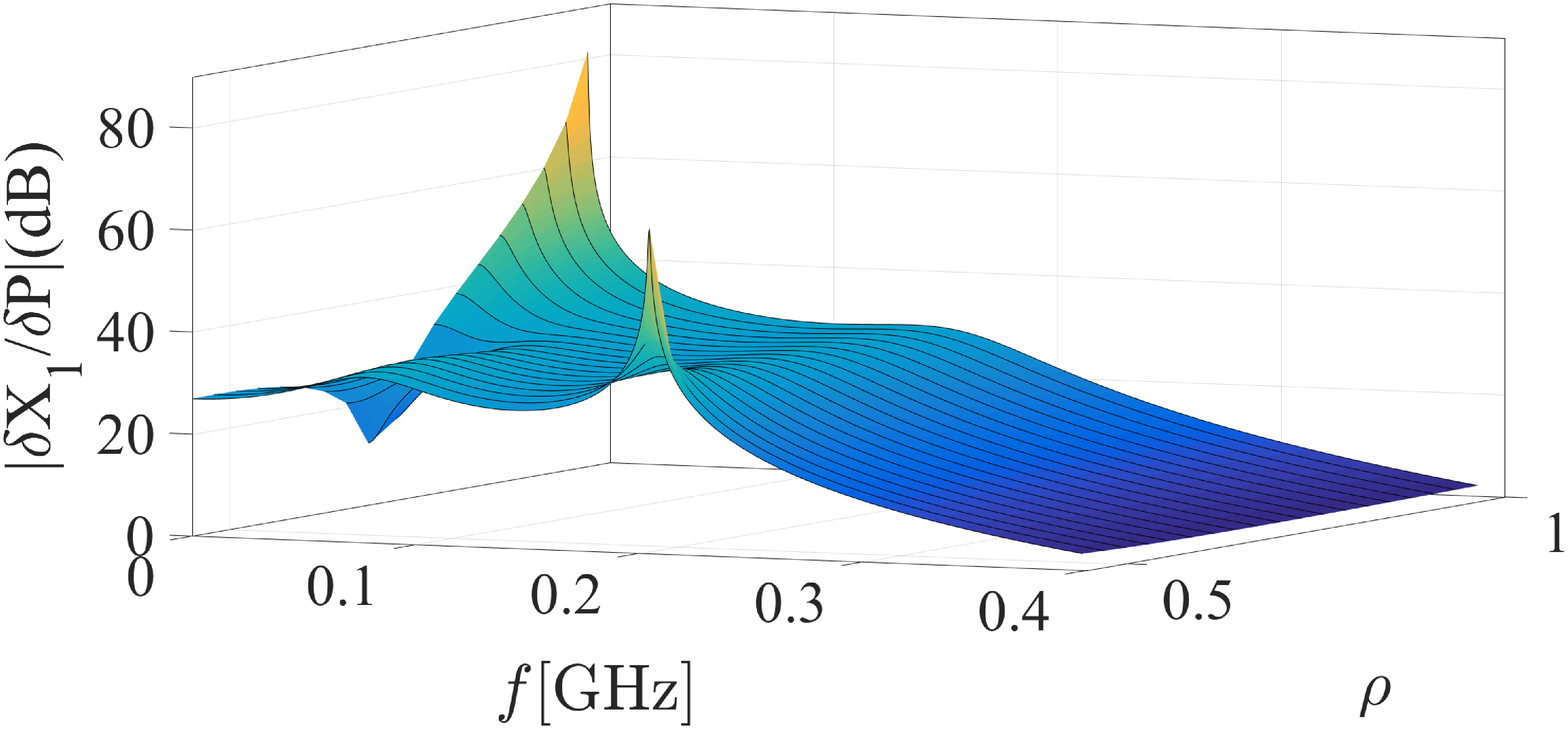}}}\\
  {\scalebox{\scl}{\includegraphics{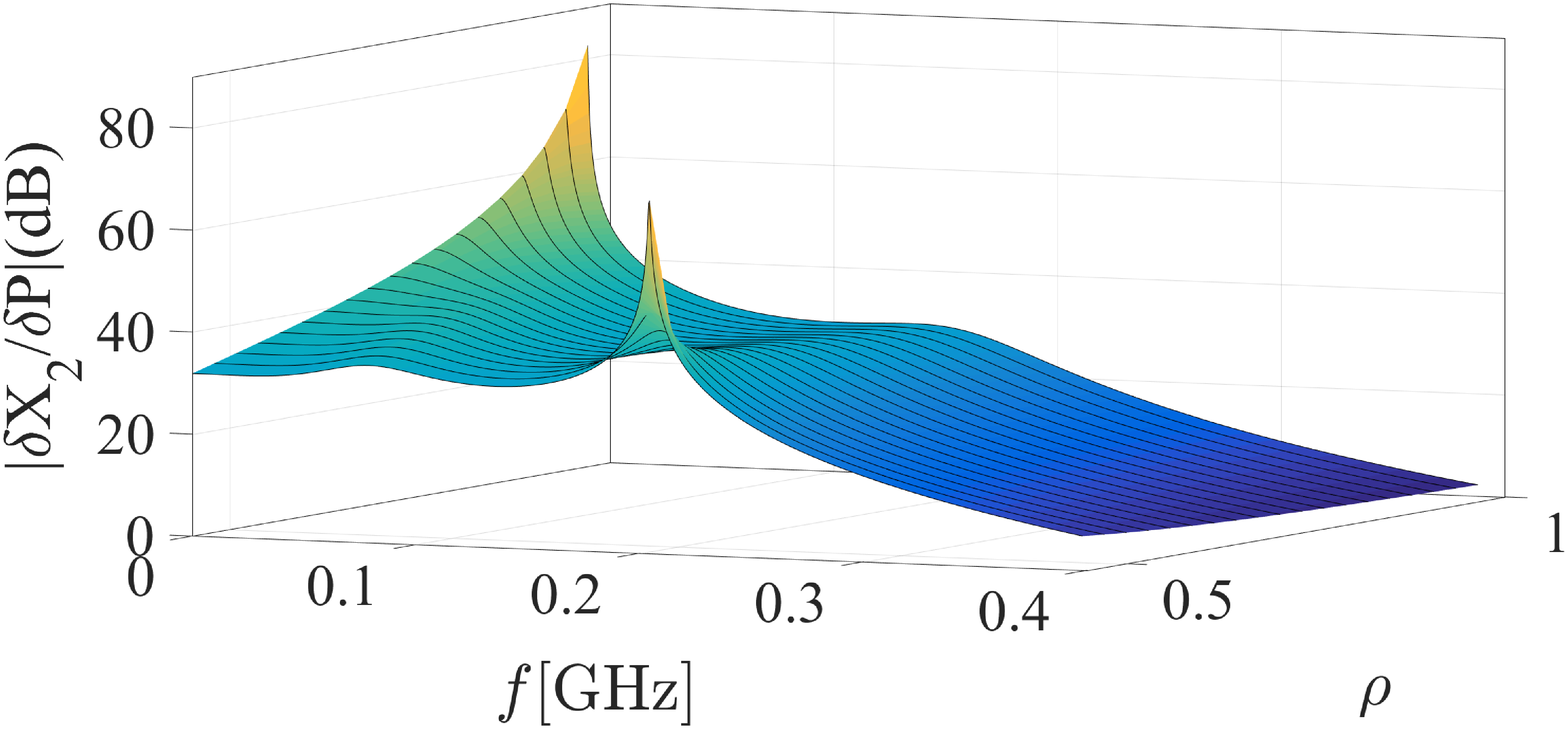}}}
  \caption{Modulation response as a function of the electric field amplitude ratio $\rho$ for $\Lambda=10^{-3}$ under out-of-phase modulation. Sharp resonances appear for strongly asymmetric phase-locked states, close to Hopf bifurcation points. The resonances appear at frequencies above the free running relaxation frequency shown as a smoother peak in the modulation response. }
  \end{center}
\end{figure}

\begin{figure}[h!]
  \begin{center}
  {\scalebox{\scl}{\includegraphics{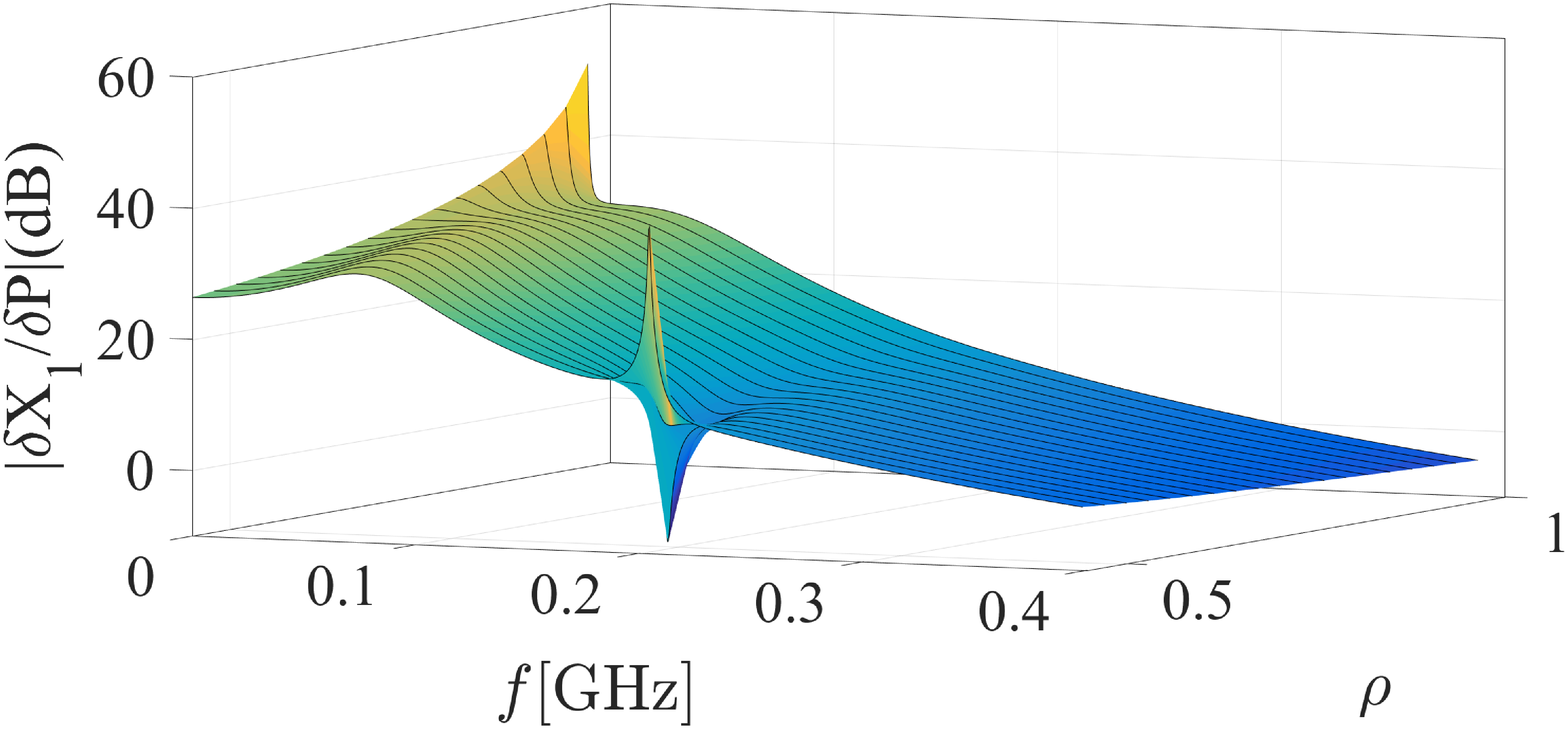}}}\\
  {\scalebox{\scl}{\includegraphics{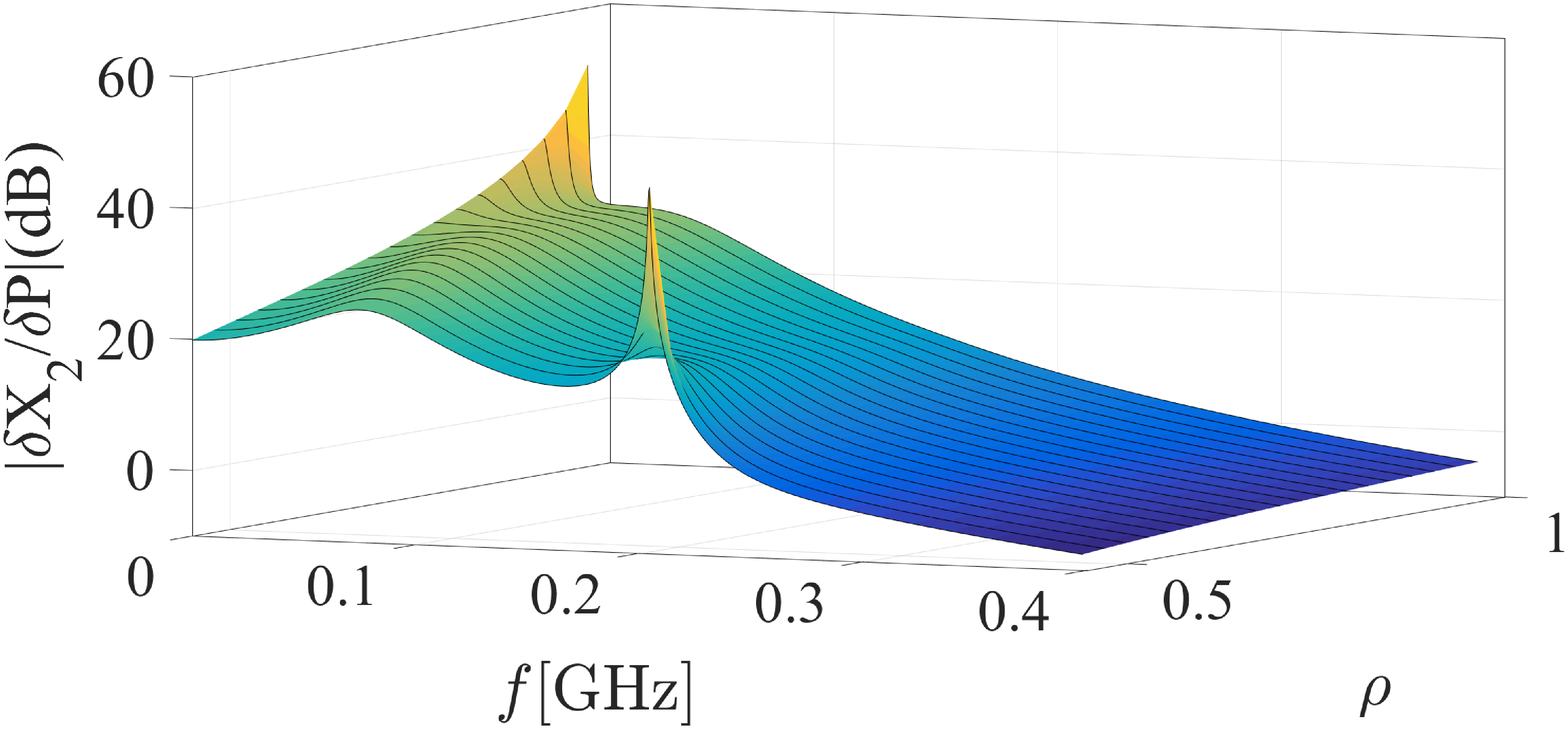}}}
   \caption{Modulation response as a function of the electric field amplitude ratio $\rho$ for $\Lambda=10^{-3}$ under in-phase modulation. Sharp resonances appear in both cases . Apart from sharp resonances, anti-resonances appear at the amplitude response of the first laser for strongly asymmetric phase-locked states, close to Hopf bifurcation points. Both resonances and anti-resonances appear at frequencies above the free running relaxation frequency shown as a smoother peak in the modulation response. }
  \end{center}
\end{figure}

\subsection{Weak coupling}
In the weak coupling regime, the stable phase-locked states can be highly asymmetric, with $\rho$ significantly different from unity, as shown in Fig. 1(a) with their free running relaxation frequencies beingat the order of 0.1\,GHz and depending on $\rho$, as illustrated in Fig. 1(b). The amplitude modulation response of the two coupled lasers for a coupling coefficient $\Lambda=10^{-3}$ is shown in Fig. 4 and 5 for the case of out-of-phase and in-phase modulation, respectively. In both case a smooth peak appears at the free running relaxation frequency and sharp peaks appear at a higher frequency for strongly asymmetric phase-locked modes. These modes are close to the stability boundary (Fig. 1(a)) where a Hopf bifurcation takes place \cite{Kominis_17a}. For the case of in-phase modulation, an interesting {\it anti-resonance} phenomenon is manifested as a sharp dip in the modulation response of the first laser, as shown in Fig. 5(a). The term anti-resonance has been used in multiple physical systems spanning the literature of cavity QED \cite{AR_1}, metamaterials \cite{AR_2} and vibration testing \cite{AR_3}. Uncharacteristically, it has never called explicitly on the photonics literature of small signal modulation of semiconductor lasers. Still a careful review of recent work \cite{AR_4} and legacy literature \cite{AR_5} shows that it was evident in multiple experimental and theoretical graphs \cite{AR_6}.

\section{Conclunding remarks}
We have considered the small signal modulation response of the simplest coupled semiconductor laser configuration consisted of a pair of twin, equally pumped lasers. The existence of symmetry-breaking modes having extended stability domains in the parameter space of the system allows for the overcoming of the restrictions in specific value ranges of the coupling coefficients when the modulation of symmetric modes is considered. Moreover, the asymmetry has been shown capable of allowing for a rich set of interesting characteristics of the modulation response such as sharp resonances and anti-resonances as well as efficient modulation at very high frequencies that are larger by orders of magnitude than the free running relaxation frequency of the system. It is worth emphasizing that these qualitative characteristics reveal the inherent features of a symmetric system under symmetry-breaking. 
Taking advantage of additional freedom in the system design by considering unequally pumped and/or detuned pairs of coupled semiconductor lasers, we can futher tailor the modulation response profile by tuning the resonances and anti-resonances at will, as well as flatenning the response and increasing the bandwidth, according to characteristics desirable for specific applications.

\section*{Acknowledgements}
This research is partly supported by two ORAU grants entitled "Taming Chimeras to Achieve the Superradiant Emitter" and ''Dissecting the Collective Dynamics of Arrays of Superconducting Circuits and Quantum Metamaterials'', funded by Nazarbayev University. Funding from MES RK state-targeted program BR05236454 is also acknowledged.

\end{document}